\documentclass[aps,prl,twocolumn,superscriptaddress]{revtex4}

\usepackage[dvips]{graphicx,color}

\input epsf.sty
\usepackage{graphicx}

\begin{document}


\title{Disappearance of antiferromagnetic spin excitations in over-doped 
La$_{2-x}$Sr$_{x}$CuO$_{4}$}

\author{S. Wakimoto}
\affiliation{ Quantum Beam Science Directorate, Japan Atomic Energy Agency,
   Tokai, Ibaraki 319-1195, Japan }

\author{K. Yamada}
\affiliation{ Institute for Materials Research, Tohoku University, Katahira,
   Sendai 980-8577, Japan }

\author{J. M. Tranquada}
\affiliation{ Brookhaven National Laboratory, Upton, New York
11973-5000, USA }

\author{C. D. Frost}
\affiliation{ ISIS Facility, Rutherford Appleton Laboratory, 
   Chilton, Didcot, OX11~0QX, UK}

\author{R. J. Birgeneau}
\affiliation{ Department of Physics, University of California, Berkeley,
   Berkeley, California 94720, USA }

\author{H. Zhang}
\affiliation{ Department of Physics, University of Toronto, Toronto,
   Ontario, Canada M5S~1A7 }

\date{\today}

\begin{abstract}

Magnetic excitations for energies up to $\sim100$~meV are studied
for over-doped  La$_{2-x}$Sr$_{x}$CuO$_{4}$ with $x=0.25$ and $0.30$,
using  time-of-flight  neutron spectroscopy.  
Comparison of spectra integrated 
over the width of an antiferromagnetic Brillouin zone demonstrates
that the magnetic scattering at intermediate energies,
$20 \lesssim \omega \lesssim 100$~meV, progressively decreases with
over-doping.
This strongly suggests that the magnetism is not related to Fermi surface 
nesting, but rather is associated with a decreasing volume fraction of 
(probably fluctuating) antiferromagnetic bubbles.

\end{abstract}


\maketitle


While remarkable progress has been made over the last few years in
identifying common features of the spin excitations in superconducting
cuprates \cite{hayd04,tran04,chri04,stoc05,rezn04}, there continues to be
considerable controversy over the microscopic origin of these dynamic
antiferromagnetic (AF) correlations.  Of particular relevance are spin
excitations with energies on the scale of the superconducting gap
\cite{woo06}.  Do these dynamic correlations reflect residual local
antiferromagnetism connected with the parent insulator (into which holes
have been doped), or, alternatively, are they a direct result of the
shape (nesting) of the electronic Fermi surface \cite{ande97}?  In the
first case, one  expects that, with sufficient doping, all vestiges of the
parent insulator should disappear, especially spin fluctuations on the
scale of the AF superexchange energy.  This is because AF correlations
are likely to survive due to charge segregation \cite{carl03}, and beyond 
some critical hole density charge segregation should become
energetically unfavorable \cite{zach02b}.  In the second case, the
existence of low-energy spin fluctuations would depend only on the
nesting properties of the Fermi surface, as for spin-density waves in
chromium and its alloys \cite{fawc94}.  The debate on this issue is
fundamental to understanding the ``normal'' state of layered cuprates,
out of which the superconductivity develops on cooling.

In this Letter, we directly test these two alternative scenarios by
experimentally searching for spin fluctuations in over-doped
La$_{2-x}$Sr$_x$CuO$_4$ (LSCO).  One recent neutron-scattering study on
over-doped LSCO~\cite{waki04} has revealed that the incommensurate
magnetic scattering at low energy ($\omega \leq 10$~meV) decreases
linearly with $T_c$ as doping increases for $x \geq 0.25$,  and disappears
coincidentally with the disappearance of the bulk superconductivity  at
$x=0.30$.  While that result suggests a clear correlation between the
magnetic excitations and superconductivity, it leaves open the
possibility that the spectral weight has shifted to higher, but still
relevant, energies.  To resolve this issue, we have used a time-of-flight
neutron-scattering technique to search for magnetic excitations at
energies up to 100 meV in over-doped LSCO with $x=0.25$ and 0.30.  
To achieve good sensitivity, we integrate the measured scattering
over wide momentum and energy intervals, and compare the results with
data for La$_{1.875}$Ba$_{0.125}$CuO$_{4}$ (LBCO 1/8) \cite{xu05}.
For $x=0.25$, we find that the magnetic scattering in the intermediate
energy  range $20\lesssim\omega\lesssim 100$~meV 
is reduced by a factor of two.  The non-superconducting
$x=0.30$ sample shows essentially no magnetic scattering for
$\omega\lesssim  60$~meV, and only weak signal for $\omega\gtrsim 60$~meV.
Given that recent angle-resolved photoemission
studies have shown that the shape of the Fermi surface becomes better
matched to a nearly-AF response in the over-doped regime
\cite{yosh06,yosh07}, opposite to the observed trend in magnetic signal
strength, the present results provide compelling evidence that
electron-hole excitations do not provide the dominant contribution to the
magnetic response in superconducting cuprates.


Single crystals of LSCO with $x=0.25$ and $0.30$ were prepared by the 
traveling-solvent floating-zone method, in the same manner as for the
crystals used for the previous triple-axis measurements
\cite{waki04}.  After growth, the crystals were  annealed in an oxygen
atmosphere to minimize any oxygen vacancies.  From magnetization 
measurements, the $x=0.25$ crystals have $T_c \sim 14$~K, while the
$x=0.30$  samples show no sign of bulk superconductivity down to 2~K,
consistent with the results of Tanabe {\it et al.} \cite{tana05}.
The crystal structures of both compositions have tetragonal ($I4/mmm$)
symmetry, with lattice constants of $a=b=3.76$~\AA~ at 10~K.  For
convenience, we will use the notation corresponding to the orthorhombic
unit cell relevant to LSCO at lower doping. In this notation, the $a$-
and $b$-axes  are parallel to  the diagonal Cu-Cu directions of the
CuO$_2$ square lattice and, therefore, the  reciprocal lattice unit
(r.l.u.) corresponds to 1.18~\AA$^{-1}$.  The orientation of the
low-energy incommensurate peaks around the AF wave vector
${\bf Q}_{\rm AF} = (1, 0)$ is indicated at the top of Fig.~2.

Neutron scattering experiments were carried out at the MAPS
time-of-flight  spectrometer installed at the ISIS pulsed neutron source,
UK.  For each composition, eight crystal rods with a total volume of $\sim
9.5$~cm$^3$ were coaligned and oriented such that the 
$[001]$ axis was parallel to the incident beam.  Measurements were 
performed either with an incident energy $E_i = 80$~meV and chopper 
frequency $f_{\rm ch}=300$~Hz or with $E_i = 140$~meV and 
$f_{\rm ch}=400$~Hz.   These cross sections are put on an absolute scale
by taking account of the sample volume and normalizing to the
signal from a standard vanadium foil.
Data have been analyzed after summing up intensities of detectors that
cover  symmetric positions in the (HK) zone to gain counting statistics.


\begin{figure}
\includegraphics[width=3.2in]{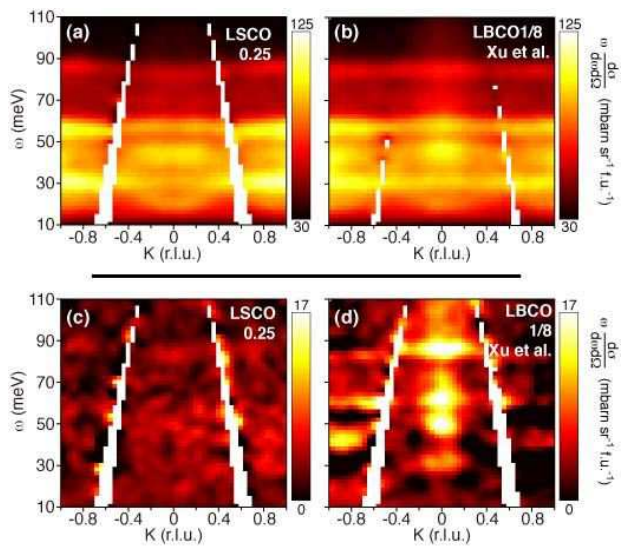}
\caption{(Color online) 
Contour plots of $\omega \frac{d \sigma}{d \Omega d \omega}$ for 
(a) LSCO $x=0.25$ and (b) LBCO $x=1/8$, all measured with $E_i=140$~meV 
at 10 K.  Data for LBCO $x=1/8$  are from Xu {\it et al.} \cite{xu05}.  
The plotted quantities are integrated for $0.5 \leq H \leq 1.5$~r.l.u.
and projected on the $K$-axis.   (c) and (d) show contour plots of
differential intensities derived by subtracting data of LSCO $x=0.30$
from those of  LSCO $x=0.25$ and those of LBCO $x=1/8$, respectively.
The white regions arising from $K \sim \pm 0.7$ correspond to areas 
without detector coverage.
}
\end{figure}

The challenge in analyzing the data is to identify the magnetic signal
in the over-doped samples.  Figures~1(a) and (b) show intensity maps of
scattering cross sections for LSCO $x=0.25$ and LBCO 1/8 (from
Xu {\it et al.} \cite{xu05}), respectively, that are integrated over the
$q$-range of $0.5\leq H\leq 1.5$~r.l.u. and then  multiplied by the
energy transfer $\omega$.  (Data for $x=0.30$ are not shown here as they
are virtually indistinguishable from $x=0.25$ in this format \cite{waki06}.)
The integration over a wide band in $H$ is done to sum up magnetic signal
that might be spread over a significant range in {\bf Q}.  The
multiplication by
$\omega$ is intended to compensate for  the energy dependence of the
scattering signals; this approach has been used previously to display the
spin-wave dispersion in stripe-ordered La$_{1.67}$Sr$_{0.33}$NiO$_4$
\cite{woo05}. In the latter case, the magnetic signal is comparable to
the phonon cross section.  In the present case, one can detect for LBCO
in Fig.~1(b) a vertical streak of magnetic signal centered around ${\bf
Q}_{\rm AF}$ ($K = 0$) superimposed on a much stronger signal from
phonons.  In contrast, for LSCO $x=0.25$ in Fig.~1(a), the signal is
entirely dominated by the phonons.
In Figs.~1(c) and (d), we have reduced most of the phonon contribution by
subtracting the data for LSCO $x=0.30$.  (Here and in Fig.~2 the LBCO
intensities have been scaled by 0.93 so that the average phonon
contribution is equal to that in the LSCO samples.) These figures show
qualitatively that the magnetic scattering that is  concentrated near
${\bf Q}_{\rm AF}$ in LBCO 1/8 has decreased significantly in LSCO $x=0.25$, 
making quantitative analysis difficult.

\begin{figure}
\includegraphics[width=3.2in]{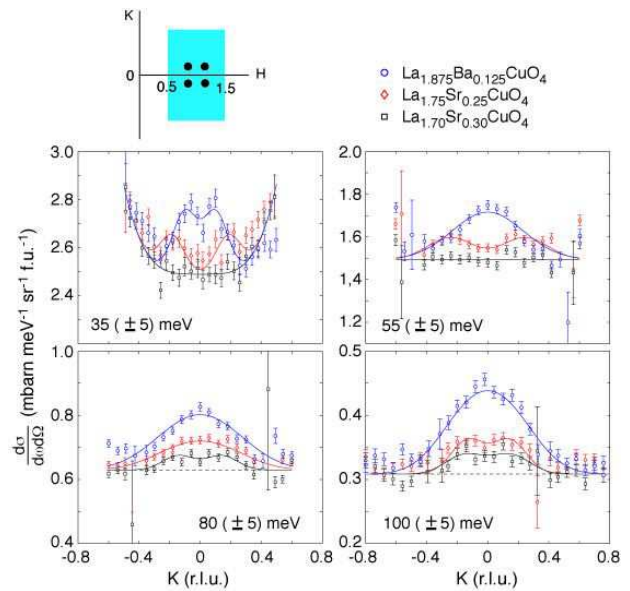}
\caption{(Color online)
Intensity profiles cut along the $K$ axis, with data integrated over 
$0.5\leq H \leq 1.5$~r.l.u., corresponding to the $H$-width of the shaded
area in  the top figure; also integrated over an energy range of $\pm5$
meV about the excitation energy specified in each panel.  All data are
taken with $E_i=140$~meV.  Circles, diamonds, and squares represent LBCO
1/8, LSCO $x=0.25$ and $0.30$, respectively.  Solid lines
are results of fits to a double-Gaussian function. 
The fits for $x=0.30$ at $\omega=35$ and 55 meV are set equal to
background; in the bottom panels, background is shown by dashed lines.
}
\end{figure}

We have found that we can identify a finite magnetic signal by taking
cuts through the data along ${\bf Q}=(1,K)$, and integrating over energy
($\omega \pm 5$ meV) as well as $H$ 
($0.5 \leq H \leq 1.5$~r.l.u).  Figure 2 shows a selection of such
intensity profiles for all three samples at four different 
energies.  (Note that for 100~meV, the $H$ integration is limited to $0.5
\leq  H \leq 1.2$~r.l.u. due to limitations in detector coverage at that
energy.)
We expect the phonon background to be very similar for all three samples,
and we take it to be a constant, except at $\omega=35$~meV where the
background has been fit to the functional form $a+bK^4$.  The {\bf
Q}-dependent signal above the background is taken to be magnetic.  (The
maximum phonon energy is 85 meV \cite{pint05}, so the signal at 100 meV
must be magnetic.)  These profiles demonstrate a monotonic decrease in the
{\bf Q}-dependent magnetic signal with doping in the intermediate energy
range, just as previously seen in the low energy regime
\cite{waki04}.

We are interested in evaluating the {\bf Q}-integrated dynamic structure
factor $S(\omega)$.  To convert from units of the differential cross
section, we make use of
\begin{equation}
\frac{d \sigma}{d \Omega d \omega} = \left(\frac{\gamma r_0}{2}\right)^2
\frac{k_f}{k_i}  f^2({\bf Q}) S({\bf Q}, \omega),
\end{equation}
where 
$\gamma r_0 / 2$ is the neutron magnetic scattering length, $k_i$ and
$k_f$  are the wave numbers of the incident and final neutrons,
respectively, and $f({\bf Q})$ is the magnetic form factor.
We have fit the cross-section profiles in Fig.~2 with a double Gaussian 
without convoluting the instrumental resolution.
Since the profiles are already integrated over the relevant range in $H$,
a simple one-dimensional integration of the fits along $K$ is
sufficient to estimate $S(\omega)$.  (The same integration range was used
for LBCO 1/8 in \cite{tran04}.) For LSCO
$x=0.30$, there is no obvious structure in the cuts at 35 and 55 meV, so
we took the data to define the background.

\begin{figure}
\includegraphics[width=3.2in]{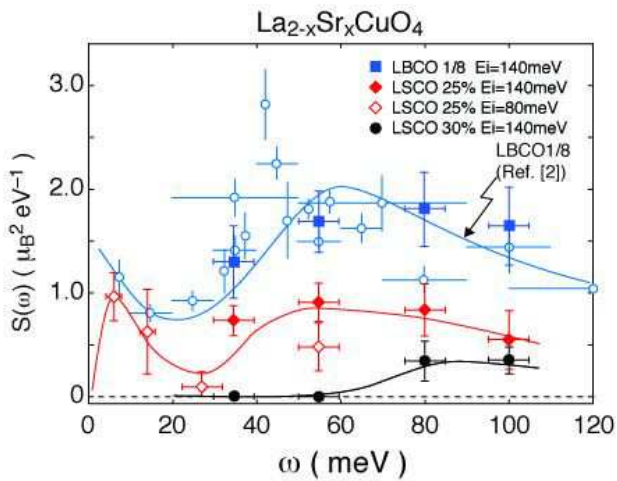}
\caption{(Color online) 
{\bf Q}-integrated dynamic structure factor $S(\omega)$ which is derived 
from the wide-$H$ integrated profiles for LBCO 1/8 (squares), LSCO 
$x=0.25$ (diamonds; filled for $E_i=140$ meV, open for $E_i=80$~meV), and
$x=0.30$ (filled circles) plotted over 
$S(\omega)$ for LBCO 1/8 (open circles) from \cite{tran04}.  
The solid lines following data of LSCO $x=0.25$ and $0.30$ are guides to 
the eyes.
}
\end{figure}

The derived $S(\omega)$ at intermediate energies for the three 
samples are shown in Fig.~3 by filled symbols.  Open diamonds represent 
results for LSCO $x=0.25$ evaluated in a similar fashion (with a flat
background) from the $E_i = 80$~meV data; the better energy resolution
with the smaller $E_i$ is beneficial for evaluating the lower energy
regime, $\omega < 30$~meV.   
The previously reported results \cite{tran04} for $S(\omega)$ in LBCO 1/8
are shown by open circles in Fig.~3 as a reference.  The results
evaluated from the new measurements \cite{xu05} for LBCO 1/8 obtained
under the same conditions as for LSCO, shown by squares, are in good
agreement.

To appreciate the significance of Fig.~3, we first note that the
magnitude of $S(\omega)$ for LBCO 1/8 over the studied energy range is
quite comparable to that recently reported for optimally doped LSCO
$x=0.16$ by Vignolle {\it et al.} \cite{vign07}.  (The results in the
latter case are reported in terms of the dynamic susceptibility,
$\chi(\omega)$, but that is equivalent to $S(\omega)$ at the low
temperatures used in these studies.)  These results are both roughly
comparable to the scattering weight found in antiferromagnetic
La$_2$CuO$_4$ \cite{hayd96a}.  Within this context, the decrease in
magnetic signal by a factor of two in LSCO $x=0.25$ is a large effect,
and the further decrease for $x=0.30$ is enormous.  We are unable to
identify any magnetic signal for $x=0.30$ at energies less than 60 meV,
which covers the energy scale relevant to superconductivity.  The
magnetic response for
$x=0.30$ is strongly depressed across an energy range characteristic of
antiferromagnetic spin fluctuations.

One common theoretical approach is to attribute the magnetic
susceptibility of the CuO$_2$ layers to electronic excitations across
the Fermi surface \cite{bulu93,dahm95,kao00}.   ARPES studies have
shown that the nested (flat and parallel) portions of the Fermi surface in
LSCO (the ones most important for the magnetic response \cite{norm07}) 
are enhanced for $x>0.15$ (see Fig.~5(f) in \cite{yosh06}). 
Furthermore, the electronic dispersion in the tetragonal [110] direction
is independent of doping for energies up to 70 meV \cite{zhou03}.  Thus,
if electron-hole excitations associated with nesting play an important
role in the dynamic susceptibility at energies up to 70 meV, we would
expect to detect  a significant response in the vicinity of ${\bf Q}_{\rm
AF}$ in our over-doped samples.  (Note that the effect of interactions,
typically included through the random phase approximation, can
redistribute weight in {\bf Q} and $\omega$ but cannot create spectral weight
where none would exist in a non-interacting system.)  The absence of such
a response in our $x=0.30$ sample and the relative weakness of the signal
for $x=0.25$ have strongly negative implications for the significance of
nesting effects.  Even if we have missed a weak magnetic signal spread
broadly in {\bf Q}, that would still be distinct from the strongly 
{\bf Q} dependent response expected from nesting (e.g., see Fig.~17 in
\cite{yosh07}).  We look to the theory community for a quantitative
analysis of the expected electron-hole contribution in the over-doped
regime.

A plausible explanation of the decrease of the total magnetic spectral 
weight is that the volume fraction of regions supporting
magnetic correlations decreases at high doping.  We have in mind that
these (probably dynamic) regions are characterized by stripe correlations
\cite{kive03,zaan01,mach89}.  In the under-doped regime, 
the incommensurability $\delta$ increases linearly with $x$, consistent
with a gradual increase in stripe density. The incommensurability
saturates for $x \gtrsim 1/8$ \cite{yama98a}, so that the stripe density
no longer increases.  A likely possibility is that, for $x>1/8$, there is
a phase separation between striped regions and more homogeneous regions.

There are independent experimental indications of phase separation in
the over-doped cuprates based on muon spin rotation studies \cite{uemu01}
and  magnetization measurements~\cite{tana05,wen02}.  There is also
evidence from scanning tunneling microscopy on
Bi$_2$Sr$_2$CaCu$_2$O$_{8+\delta}$ for large variations in the local
magnitude of the superconducting gap \cite{fang06,jame06}. A decrease of
the superconducting volume fraction has been reported to occur in LSCO for
$x\geq 0.20$ \cite{uemu01,tana05,wen02}; note that this is considerably
higher than the point $x=1/8$, above which the CuO$_2$ planes cannot be
uniformly stripe correlated.  If one associates the superconductivity
with the striped regions \cite{emer97}, then we speculate that the
decrease of the superconducting volume fraction begins when the
separation between neighboring striped regions exceeds the
superconducting coherence length \cite{kivepc}.

In conclusion, we have shown that, in over-doped LSCO, spin excitations
effectively decrease with doping and disappear at $x=0.30$
for energies typical of antiferromagnetic fluctuations.  We have
argued that this is strong evidence that the magnetic signal in optimally
doped cuprates cannot have a dominant contribution from conventional
electron-hole excitations.  Instead, the antiferromagnetic spin
correlations in superconducting samples must be vestiges of the parent
insulator.

The authors gratefully acknowledge G. Xu for sharing the data of LBCO 1/8
prior to  publication.  We also thank K. Kakurai, H.-K. Kim, A.
Kagedan, and S. A. Kivelson for invaluable  discussions.  This work is
partially supported by the Japan-UK Collaboration  Program on Neutron
Scattering.  SW is supported by a Grant-In-Aid for Young  Scientists B
from the Japanese Ministry of Education, Culture, Sports, Science and 
Technology.  JMT is supported at Brookhaven by the Office of Science,
U.S. Dept.  of Energy, under Contract No. DE-AC02-98CH10886.  RJB is
supported at Lawrence  Berkeley Laboratory by the Office of Basic Energy
Sciences, U.S. Dept. of Energy  under contract No. DE-AC03-76SF00098.


\end{document}